\documentclass[aps,prc,twocolumn,superscriptaddress,floatfix,amsmath,amssymb,nofootinbib,longbibliography]{revtex4-2}

\usepackage{mathtools} 
\usepackage{amsfonts}
\usepackage{amsthm}
\usepackage{mleftright} \mleftright 
\usepackage[dvipsnames]{xcolor}
\usepackage{xspace}
\xspaceaddexceptions{]\}}
\usepackage{bm} 
\usepackage{braket}
\usepackage{array} 
\usepackage{booktabs} 
\usepackage[caption=false]{subfig}
\usepackage{orcidlink}
\usepackage{hyperref}
\hypersetup{
  colorlinks   = true,
  urlcolor     = blue, 
  linkcolor    = blue, 
  citecolor   = blue 
}
\usepackage{cleveref}

\crefname{table}{Tab.}{Tabs.}
\Crefname{table}{Table}{Tables}
\crefname{figure}{Fig.}{Figs.}
\Crefname{figure}{Figure}{Figures}
\crefname{equation}{Eq.}{Eqs.}
\Crefname{equation}{Equation}{Equations}
\crefname{section}{Sec.}{Secs.}
\Crefname{section}{Section}{Sections}
\crefname{appendix}{Appendix}{Appendices}

\newcommand{\eg}{e.g.}
\newcommand{\ie}{i.e.}
\newcommand{\cf}{cf.}
\newcommand{\ai}{ab initio\xspace}

\newcommand{\leftd}{\left.\kern-\nulldelimiterspace}
\newcommand{\rightd}{\right.\kern-\nulldelimiterspace}

\newcommand{\unit}[1]{\ \text{#1}} 
\newcommand{\MeV}{\unit{MeV}}

\newcommand{\rcite}[1]{Ref.~\cite{#1}}
\newcommand{\rscite}[1]{Refs.~\cite{#1}}

\newcommand{\freq}{{\hbar \omega}}
\newcommand{\emax}{e_\text{max}}
\newcommand{\emaxmax}{\emax^\text{max}}
\newcommand{\emaxComp}[1]{\emax^{(#1)}}
\newcommand{\emaxBalancing}{e_\text{bal}}
\newcommand{\emaxBalancingComp}[1]{\emaxBalancing^{(#1)}}
\newcommand{\emaxBalancingHF}{\emaxBalancing^\text{HF}}
\newcommand{\emaxvec}{\mathbf{\emax}}
\newcommand{\emaxBalancingvec}{\mathbf{\emaxBalancing}}
\newcommand{\eThreeMax}{e_{3\text{max}}}
\newcommand{\ndim}{n_\text{dim}}

\newcommand{\OrderRatio}[1]{R^{(#1)}}
\newcommand{\EnergyCorrection}[1]{E^{(#1)}}
\newcommand{\MBPTEnergy}[1]{E_{\text{MBPT}(#1)}}
\newcommand{\CCOrderRatio}[1]{R^\text{#1}}
\newcommand{\HFEnergy}{E_\text{HF}}

\newcommand{\elm}[2]{$^{#2}$#1} 

\newcommand{\MagicInt}{1.8/2.0\,(EM)\xspace}
\newcommand{\DoublyMagicInt}{1.8/2.0\,(EM7.5)\xspace}

\newcommand{\Error}{\sigma}
\newcommand{\PracticalError}{\widetilde\sigma}

\newcommand{\MBError}{\Error^\text{MB}}
\newcommand{\AcademicMBError}[1]{\MBError_{\text{MBPT}({#1})}}
\newcommand{\PraMBError}{\PracticalError^\text{MB}}
\newcommand{\PracticalMBError}[1]{\PraMBError_{\text{MBPT}({#1})}}

\newcommand{\MSError}{\Error^\text{basis}}
\newcommand{\AcademicMSError}[1]{\MSError_{\text{MBPT}({#1})}}
\newcommand{\AcademicMSErrorComp}[1]{\MSError_{({#1})}}
\newcommand{\PraMSError}{\PracticalError^\text{basis}}
\newcommand{\PracticalMSError}[1]{\PraMSError_{\text{MBPT}({#1})}}
\newcommand{\PracticalMSErrorComp}[1]{\PraMSError_{({#1})}}

\newcommand{\ExactEnergy}{E_\infty}

\newcommand{\MostConvergedEnergy}[1]{\MBPTEnergy{#1}(\emaxmax)}

\makeatletter
\AtBeginDocument{\let\LS@rot\@undefined}
\makeatother

\begin{document}

\title{Balancing theory uncertainties in ab initio nuclear structure calculations: \\
Many-body truncation versus finite basis size}

\author{L.~Zurek \orcidlink{0000-0003-0763-7613}}
\email{lars.zurek@cea.fr} 
\affiliation{CEA, DAM, DIF, 91297 Arpajon, France}
\affiliation{Université Paris-Saclay, CEA, Laboratoire Matière en Conditions Extrêmes, 91680 Bruyères-le-Châtel, France}

\author{U.~Vernik \orcidlink{0000-0002-3717-945X}}
\email{urban.vernik@tu-darmstadt.de}
\affiliation{Technische Universit\"at Darmstadt, Department of Physics, 64289 Darmstadt, Germany}
\affiliation{ExtreMe Matter Institute EMMI, GSI Helmholtzzentrum f\"ur Schwerionenforschung GmbH, 64291 Darmstadt, Germany}

\author{P.~Demol \orcidlink{0000-0003-2511-7179}}
\email{pepijn.demol@ulb.be} 
\affiliation{Universit\'e Libre de Bruxelles, Institut d’Astronomie et d’Astrophysique, 1050 Brussels, Belgium}
\affiliation{
Brussels Laboratory of the Universe -- BLU-ULB, 1050 Brussels, Belgium}

\author{T.~Duguet \orcidlink{0000-0002-7596-3851}}
\email{thomas.duguet@cea.fr} 
\affiliation{IRFU, CEA, Universit\'e Paris-Saclay, 91191 Gif-sur-Yvette, France}

\author{M.~Frosini \orcidlink{0000-0002-8330-4004}}
\email{mikael.frosini@cea.fr} 
\affiliation{CEA, DES, IRESNE, DER, SPRC, LEPh, 13108 Saint-Paul-lez-Durance, France}

\author{A.~Tichai \orcidlink{0000-0002-0618-0685}}
\email{alexander.tichai@tu-darmstadt.de}
\affiliation{Technische Universit\"at Darmstadt, Department of Physics, 64289 Darmstadt, Germany}
\affiliation{ExtreMe Matter Institute EMMI, GSI Helmholtzzentrum f\"ur Schwerionenforschung GmbH, 64291 Darmstadt, Germany}
\affiliation{Max-Planck-Institut f\"ur Kernphysik, 69117 Heidelberg, Germany}

\date{August 19, 2026}

\begin{abstract}
First-principles calculations of atomic nuclei are necessarily incomplete as the Schr\"odinger equation is solved using approximate methods and due to the finite dimension of the employed Hilbert space.
By balancing many-body truncation and basis-size uncertainties, we formalize a criterion for the optimal one-body basis dimension in a given ab initio nuclear structure computation.
Next, it is demonstrated that higher-order many-body contributions can be computed using smaller basis sizes than used for the lower orders when a consistent accuracy in the calculation is targeted.
Our findings are empirically validated using many-body perturbation theory and coupled-cluster calculations of nuclei spanning a large portion of the nuclear chart using two sets of chiral two- and three-nucleon interactions.
The results suggest that considerable computational savings can be obtained using many-body-order-dependent one-body basis sizes.
\end{abstract}

\maketitle

\section{Introduction}\label{sec:intro}

The essence of the present work can be summarized as follows: \emph{``There is no point in working hard to reduce a small uncertainty when a larger uncertainty is present"}.
This simple statement has profound implications for scientific research.
It is crucial for experimental design~\cite{Drosg2007Uncertainties,Coleman2018ExperimentationUncertainties,Melendez2021ComptonScatteringExpDesign},
and should guide theorists when developing methods and performing calculations.
In many applications, uncertainties from nuclear theory are among the largest at play, \eg{}, in the search for neutrinoless double beta decay~\cite{Agostini2023NeutrinolessDoubleBeta}, neutrino flux measurements~\cite{DUNE2023CrossSectionUncertainties}, and astrophysical calculations~\cite{Kullmann2022NuclearUncertaintiesMergers}.
Similarly, experimental errors on nuclear observables are often smaller in size than theoretical uncertainties~\cite{wang_ame_2021, angeli_TableExperimental_2013, moller_NuclearGroundstate_2016, Grams2026BSkG5, Stroberg2021, Miyagi2025AbInitioRadiiReview, Wang2026LatticeEFTMoments}. Consequently, one of the key objectives today is to improve the accuracy of nuclear theory predictions.

The \ai{} approach to atomic nuclei with its promise of being systematically improvable~\cite{Ekstroem2023abinitio} is suitable for this.
Given the extensive advances in \ai{} nuclear theory, nuclear properties can be calculated over a significant portion of the Segrè chart using correlation-expansion methods~\cite{Arthuis2026DoublyMagic, Tichai2024BCCCaNiSn, Miyagi2022HeavyAbInitio, Hu2022AbInitio208Pb, Demol:2026glc, Vernik2026}, thanks to their mild computational scaling with system size.
Still, computational complexity increases steeply as the many-body truncation is relaxed, rendering accurate calculations a formidable challenge. The cost of a calculation with a correlation-expansion method scales polynomially like $\mathcal{O}(\ndim^k)$ with respect to the dimension of the one-body basis $\ndim$.
The order of the many-body truncation is reflected in the exponent $k$, with higher orders yielding larger exponents. 
Thus, the computational cost increases dramatically when pushing towards high-accuracy calculations~\cite{Binder2013CCwithFull3N,Heinz2020,Heinz2025, Li2026HighOrderMBPT}, to heavy nuclei~\cite{Hu2022AbInitio208Pb, Tichai2024BCCCaNiSn,Bonaiti2025heavy,Bonaiti2026sn140}, to open-shell systems~\cite{Soma:2012zd,Tichai:2018vjc,Novario2020a,Frosini22b,Frosini22c,Yuan2022,Marino2026CCGroundStates, Demol:2026glc} or to all three at once~\cite{Vernik2026}. 
While the first aspect requires computations at larger many-body orders, the last three effectively require larger and larger one-body bases.

In this context, it is highly desirable to optimize $\ndim$ given a certain value of $k$. 
With the basic principle stated in the second sentence of this introduction in mind, we characterize precisely how small the basis can be chosen by comparing sizes of different uncertainties. 
Although the ideas presented here apply to all sources of theoretical uncertainties~\cite{Hergert2020AbInitio, Ekstroem2023abinitio}, the present focus is on balancing two types of uncertainties: those stemming from the truncation of the many-body expansion at a given working order (many-body truncation uncertainty) and those arising from using a finite one-body basis (basis-size uncertainty).
Doing so, a criterion is formalized to identify an ``optimal'' one-body basis size for a given calculation of interest such that the intrinsic accuracy of the many-body calculation can be realized at the lowest possible computational cost.
For simplicity, the analysis is restricted to ground-state energies of representative nuclei and is based mainly on many-body perturbation theory calculations.
We also elaborate on the necessary steps to extend the principles distilled here to other nuclei, Hamiltonians, correlation-expansion methods, and observables.

In a second step, we prove that even more optimal numerical calculations can be achieved by adapting the one-body basis dimension to the contribution arising at each order in the many-body expansion. This is of great interest to avoid wasting finite resources and is becoming increasingly more important with the advent of more sophisticated \ai calculations~\cite{He2024imsrg3f2,Heinz2025,Vernik2026}.
In particular for large-scale systematic studies, which are becoming possible within the frame of \ai nuclear theory~\cite{Stroberg2021}, good resource management is paramount.
This work offers a pragmatic route to optimize the computational cost in \ai{} calculations of atomic nuclei.

The paper is organized as follows. 
In \cref{sec:academic_analysis}, we discuss theory uncertainties and how balancing them leads to a one-body basis of optimal size.
In \cref{sec:mix}, a simple prescription to reduce the cost of correlation-expansion calculations by computing different contributions with different one-body-basis dimensions is introduced.
An extensive numerical benchmark is presented in \cref{sec:chart}, which is extended to coupled-cluster calculations in \cref{sec:CC}. We conclude with a list of practical recommendations in \cref{sec:rec} and summarize our findings in \cref{sec:summary}.

\section{Theory uncertainties}\label{sec:academic_analysis}

The present work focuses on theoretical uncertainties originating from many-body truncations and the use of a finite basis to represent operators and the many-body wavefunction.
Systematic uncertainties associated with the construction of the input Hamiltonian at a given order in the chiral expansion and statistical uncertainties arising from the adjustment of low-energy constants to experimental data ~\cite{Furnstahl2015Uncertainties,Melendez2019Uncertainties,Hu2022AbInitio208Pb,Plies2025UncertaintiesSVD, Millican2026ChiralConvergence} are not discussed. 
While interaction uncertainties may dominate, their proper assessment requires extensive survey studies with explicit calculations at various orders in the chiral power counting. This is beyond the scope of this study.

\subsection{Uncertainties from many-body truncation}
\label{sec:mb}

In \ai{} nuclear structure calculations, the many-body Schrödinger equation is solved using an (intrinsic) Hamiltonian.
For mid-mass and heavy nuclei, this is commonly achieved using correlation-expansion methods. 
Starting from a reference state $\ket{\Phi}$, dynamical correlations in the wavefunction are accounted for through a systematic expansion in terms of particle-hole excitations on top of this reference state.
Going to higher orders in this expansion, the wavefunction is expected to converge to the exact solution.
The reader is referred to \rcite{Hergert2020AbInitio} for an overview of many-body methods in \ai{} nuclear theory.

In the first part of this work we employ many-body perturbation theory (MBPT), where the ground-state energy is expanded as a formal power series 
\begin{align}\label{eq:FormalMBPT}
    \ExactEnergy(\lambda) \equiv \sum_{p=0}^\infty \lambda^p \EnergyCorrection{p} 
\end{align}
in terms of an auxiliary parameter $\lambda$, which is introduced through the splitting of the Hamiltonian according to $H \equiv H_0 + \lambda H_1$. 
Eventually, one sets $\lambda=1$ to obtain the ground-state energy $\ExactEnergy$ of interest.
The unperturbed part of the Hamiltonian $H_0$ is chosen such that it can be solved exactly with the reference state $\ket{\Phi}$ as its ground state.
The total energy of the reference state is given by
\begin{equation}\label{eq:ReferenceEnergy}
    E_\text{ref} \equiv \EnergyCorrection{0} + \EnergyCorrection{1} \,.
\end{equation}
In the present work, $\ket{\Phi}$ is taken to be a canonical Hartree-Fock (HF) state such that $E_\text{ref} = \HFEnergy$. The energy corrections at \emph{orders} $p>1$ correspond to (dynamical) particle-hole correlations induced by the perturbation operator $H_1$.
Their practical evaluation in many-body calculations is based on a diagrammatic formulation building on Wick's theorem.
The reader is referred to \rcite{Tichai2020MBPTReview} for a review of MBPT.

Due to the strong increase in cost with increasing order, \cref{eq:FormalMBPT} must be truncated in practice. 
Restricting the sum to $p \leqslant P$ yields the MBPT($P$) approximation of the ground-state energy,
\begin{equation}\label{eq:TruncatedMBPT}
    \MBPTEnergy{P} = \HFEnergy +  \sum_{p=2}^P \EnergyCorrection{p} \,.
\end{equation}
We call $P$ the \emph{truncation order} and refer to $\HFEnergy, \EnergyCorrection{2}, \EnergyCorrection{3}, \dots$ as \emph{contributions} from different orders.

At a finite truncation order $P$, the many-body truncation error of the ground-state energy associated with the Hamiltonian under consideration is given by 
\begin{equation}
    |\MBPTEnergy{P} - \ExactEnergy| \,.
\end{equation}
As an exact solution is not available beyond the lightest systems~\cite{Roth2007ITNCSM}, the dependence of the error on $\ExactEnergy$ is removed by assuming that the error of a calculation at truncation order $P$ is dominated by the leading discarded contribution $\EnergyCorrection{P+1}$.
This is in line with the empirically observed convergence patterns of high-order calculations of light nuclei when using a soft chiral interaction~\cite{Tichai2016HighOrderMBPT}.
Thus, a simple estimate for the many-body truncation (MB) uncertainty of the ground-state energy computed at order $P$ reads
\begin{equation}\label{eq:BasicMBError}
    \AcademicMBError{P} \equiv |\MBPTEnergy{P} - \MBPTEnergy{P+1}| = |\EnergyCorrection{P+1}|.
\end{equation}

Later, the uncertainty of MBPT(2) calculations is evaluated by computing \cref{eq:BasicMBError} explicitly.
To this end, we employ a one-body basis of large dimension characterized by $\emaxmax$, which we introduce in \cref{sec:ms}.
Given that calculations beyond third order are not reported in this work, the many-body truncation uncertainty of MBPT(3) calculations needs to be estimated in a different way.
To this end, \cref{eq:BasicMBError} is rewritten as
\begin{equation}\label{eq:MBError}
    \AcademicMBError{P} = \OrderRatio{P} |\EnergyCorrection{P}| \, ,
\end{equation}
where
\begin{equation}
  \OrderRatio{P} \equiv \left|\frac{\EnergyCorrection{P+1}}{\EnergyCorrection{P}}\right| \,.
\end{equation}
Employing \cref{eq:MBError} relies on estimating $\OrderRatio{P}$.
This is achieved based on the expectation that $\OrderRatio{P}$ is similar in all nuclei, which follows from the size-extensive character of the Rayleigh-Schrödinger formulation of MBPT~\cite{Nooijen2005SizeExtensivityConsistency,ShavittBartlett2009Book}. 
This has been empirically confirmed for low-order MBPT calculations over a wide range of nuclear masses~\cite{Svensson2025MBPTUncertainties}.
Small deviations from this rule are attributed to finite-size and shell effects, as well as to isospin dependence associated with different $N/Z$ ratios.
Therefore, size extensivity allows us to infer uncertainties in heavier systems based on the MBPT convergence in light and medium-mass nuclei where basis-size convergence is obtained more easily. 
For example, MBPT calculations at fourth order give $\OrderRatio{3} \approx$ 0.49 in \elm{O}{16} and 0.40 in \elm{Ca}{48}~\cite{Li2026HighOrderMBPT,LiPrivComm} when using the \MagicInt Hamiltonian~\cite{Hebeler2011Magic}. Based on these values, the mean value $\OrderRatio{3} = 0.44$ is presently chosen to estimate many-body truncation uncertainties of MBPT(3) calculations with that Hamiltonian.

An alternative to \cref{eq:MBError} was recently proposed in \rcite{Svensson2025MBPTUncertainties}.
Based on a Bayesian uncertainty model, a truncation-order-independent posterior distribution for $\OrderRatio{P}$ was inferred from MBPT(2/3) calculations, and the authors determined $\OrderRatio{P} = 0.156$ as the most likely value for the \MagicInt Hamiltonian.
The presently employed value is significantly larger as it has been determined based on selected fourth-order calculations. This results in larger many-body truncation uncertainties compared to \rcite{Svensson2025MBPTUncertainties}.

Other empirical estimates for the many-body truncation uncertainty have been proposed, \eg{}, using the spread of different many-body methods~\cite{Hergert2013IMSRG3N, Hergert2020AbInitio}, the difference between explicitly computed orders in a correlation expansion~\cite{Simonis2019Responses,Miyagi2022HeavyAbInitio,Bonaiti2026sn140}, or variations of the reference state. Although such strategies probe the sensitivity to higher-order effects, they do not constitute a sound uncertainty model. 
In any case, the balancing method proposed in the present work can be used with any many-body truncation uncertainty model, although the quality of the results will (to a certain degree) depend on it.

\subsection{Uncertainties from finite basis size}\label{sec:ms}

The second type of uncertainty considered in this work originates from the finite dimension of the one-body basis used to represent operators and many-body states.
In ab initio nuclear structure calculations, one-body basis states are commonly obtained as eigenstates of a spherically symmetric harmonic oscillator (sHO) characterized by its frequency $\freq$.
In practice, only one-body basis states up to a maximum value $\emax$ of the parameter $e = (2n+l)$ are kept.
Here, $n$ ($l$) denotes the principal (orbital angular momentum) quantum number. 
Correspondingly, the ground-state energy obtained at truncation order $P$ based on a truncated basis characterized by $\emax$ is denoted as $\MBPTEnergy{P}(\emax)$. 

The number $\ndim$ of one-body basis states, which sets the cost of a given many-body calculation, scales as $\mathcal{O}(\emax^3)$. Taking MBPT(3) as an example, the runtime of the computation is asymptotically proportional to $A^2 \ndim^4$, where $A$ is the mass number of the nucleus.
This shows why an optimal choice of $\emax$ is crucial to limit the cost of a computation.

\begin{figure}[tbp]
\includegraphics[width=.9\linewidth]{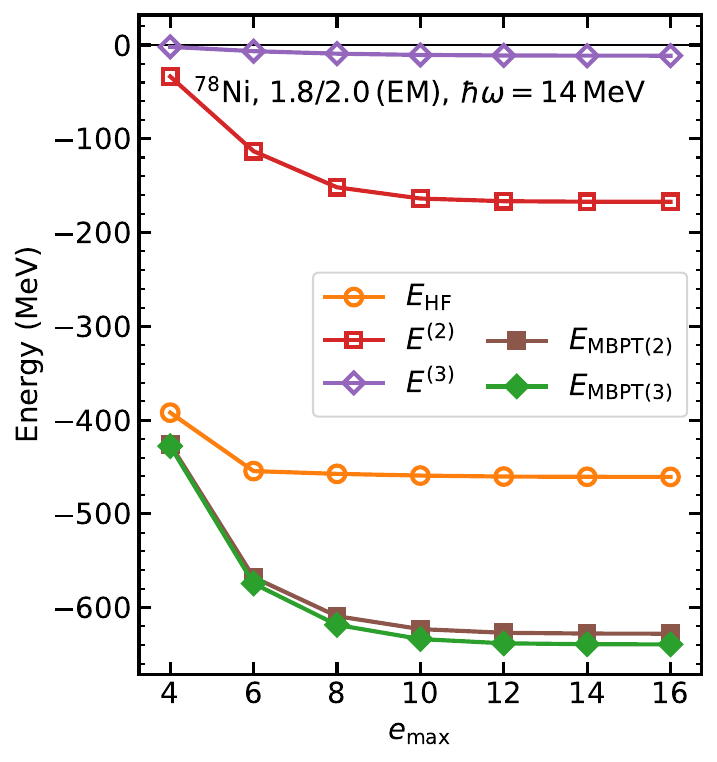}
\caption{Basis-size convergence of different MBPT corrections and partial sums for \elm{Ni}{78}. All calculations use the \MagicInt Hamiltonian and the sHO frequency $\freq = 14 \MeV$.}
\label{fig:convergence}
\end{figure}

The fact that a calculation at finite $\emax$ does not agree with the 
\emph{infinite basis-size limit} $\emax \to \infty$ is reflected in the basis-size error defined as
\begin{equation}\label{eq:InfinityMSError}
    |\MBPTEnergy{P}(\emax) - \MBPTEnergy{P}(\infty)| \,.
\end{equation}
By increasing $\emax$, the result converges towards $\MBPTEnergy{P}(\infty)$ and the basis-size error is reduced.\footnote{In this limit, all calculations are independent of the particular choice of the sHO frequency.} Such a convergence pattern is exemplified in \cref{fig:convergence}, which shows MBPT energy corrections and partial sums for the doubly-magic nucleus \elm{Ni}{78} calculated at different $\emax$.

Since  $\MBPTEnergy{P}(\infty)$ is not available in practice, the basis-size error  \cref{eq:InfinityMSError} is approximated with
\begin{align}
    \AcademicMSError{P}(\emax)
    &\equiv |\Delta \MBPTEnergy{P}(\emax)|  
    \label{eq:MSError} \\
    &\equiv |\MBPTEnergy{P}(\emax) - \MostConvergedEnergy{P}| \, , \notag
\end{align}
where $\emaxmax$ labels the largest $\emax$ actually considered, thus removing any explicit dependence on the infinite basis-size limit. This uncertainty model is based on the rapid convergence with $\emax$ that is empirically observed, thus providing a good estimate for the true error as long as $\AcademicMSError{P}(\emax)$ is much greater than the basis-size error of $\MostConvergedEnergy{P}$ itself.

\subsection{Balancing uncertainties}\label{sec:balancing}

After discussing the relevant uncertainties, the objective is to identify a balancing point in theoretical calculations to minimize computational resources while keeping the predictive power of the calculations on par with traditional approaches.

\begin{figure}[tbp]
\includegraphics[width=.9\linewidth]{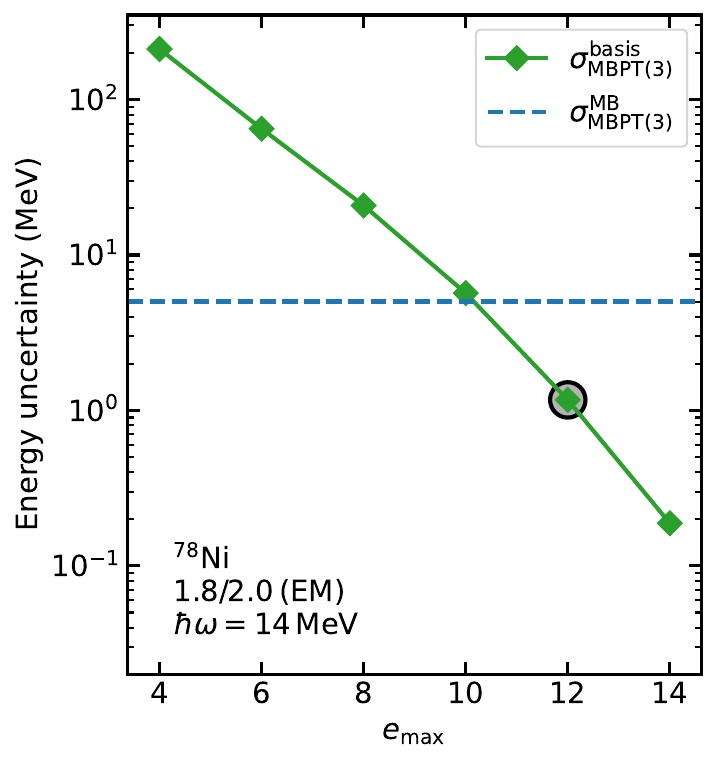}
\caption{Basis-size and many-body truncation uncertainties for the MBPT(3) calculations shown in \cref{fig:convergence}. The resulting balancing point is marked with a black circle.}
\label{fig:BasicBalancing}
\end{figure}

In \cref{fig:BasicBalancing}, many-body truncation and basis-size uncertainties of the \elm{Ni}{78} ground-state energy are displayed
based on the MBPT(3) data from \cref{fig:convergence}. 
The many-body truncation uncertainty is computed using Eq.~\eqref{eq:MBError}, evaluated with $\OrderRatio{3} = 0.44$ and $\emaxmax = 16$.
The basis-size uncertainty is obtained using Eq.~\eqref{eq:MSError} for calculations ranging from $\emax=4$ to 14.
The basis-size uncertainty is seen to decrease exponentially with $\emax$.
Beyond the critical value of $\emax = 12$, the many-body truncation uncertainty fully dominates.
Therefore, increasing $\emax$ much beyond the point where the two uncertainties are equal does not increase the accuracy of the prediction that is limited by the dominating many-body truncation uncertainty.
Hence, both uncertainties should be \emph{balanced}.
We refer to the smallest $\emax$ value such that  
\begin{equation}\label{eq:BalancingCondition}
    \AcademicMSError{P}(\emax) < \AcademicMBError{P}
\end{equation}
as the \emph{``balancing point''} and denote it by $\emaxBalancing$.
We consider this to be the optimal basis size for the many-body calculation at the given truncation order $P$ (see Ref.~\cite{Hagen2010CC} for a similar conclusion).

Let us discuss the reasons for and the consequences of this balancing procedure in more detail. 
The many-body truncation uncertainty is intrinsic to the truncation of the correlation expansion at order $P$ and characterizes its accuracy. 
This uncertainty is present even in the infinite basis-size limit.
Therefore, there is little value in decreasing the basis-size uncertainty below a threshold set by the many-body truncation uncertainty.
Instead, the basis size can be selected such that many-body truncation and basis-size uncertainties are of comparable size.
Balancing uncertainties allows one to fully exploit the accuracy of a given many-body truncation at minimal computational cost.

When increasing the truncation order, the many-body truncation uncertainty is reduced.
The increased target accuracy shifts the balancing point towards larger $\emax$ values. 
Conversely, if a reduction of the basis-size uncertainty to the size of the many-body truncation uncertainty is unachievable, the inclusion of higher-order many-body contributions is irrelevant, as the total uncertainty will remain dominated by the basis-size uncertainty in such a case.

The location of the balancing point depends in principle on the basis-size and many-body truncation uncertainty models used. 
For example, employing Eq.~\eqref{eq:MBError} with $\OrderRatio{3}<0.44$ would lead to a smaller estimate of the many-body truncation uncertainty than the one depicted in \cref{fig:BasicBalancing}.
This would push the point where the many-body truncation uncertainty matches the basis-size uncertainty to larger $\emax$ such that $\emaxBalancing$ might also be increased.
This shows that a smaller estimate of the many-body truncation uncertainty can lead to a more conservative setting of the calculation parameters.
In the present case, a large range $\OrderRatio{3} \in [0.11, 0.50]$, including the value $\OrderRatio{3}=0.156$ inferred from calculations up to third order~\cite{Svensson2025MBPTUncertainties}, is found to lead to the same balancing point as obtained here.
Rather extreme values $\OrderRatio{3} < 0.023$ or $\OrderRatio{3} > 1.8$ (suggesting a diverging MBPT series) are needed to shift the balancing points by more than 2 units.
This exemplifies the robustness of the procedure with respect to reasonable choices of the uncertainty models.

\section{Order-dependent basis sizes}\label{sec:mix}

\subsection{General rationale}\label{sec:mix_idea}

The previous section discussed how to find an optimal basis size for calculations using a correlation-expansion method at a given truncation order $P$.
So far, all contributions up to $P$ have been computed using the same one-body basis size. This traditional setup is referred to as calculations based on a \emph{uniform basis size}. This setup is however not the only possible choice. Indeed, reconsidering \cref{fig:convergence}, one observes that contributions from different orders $p\leqslant P$ in the many-body expansion converge at vastly different rates towards their infinite basis-size limits. It is possible to leverage this observation by computing contributions from different orders $p$ with one-body bases truncated to different sizes, \cf\ \rscite{Bonaiti2025heavy,Li2026HighOrderMBPT}.

\begin{figure}[tbp]
\includegraphics[width=.9\linewidth]{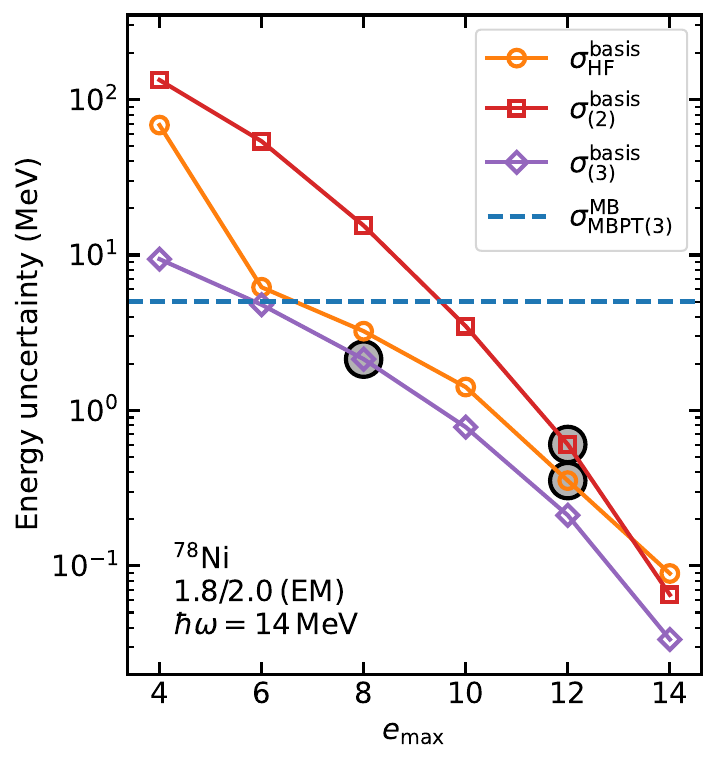}
\caption{Same as \cref{fig:BasicBalancing} with  basis-size uncertainties displayed separately for contributions originating from different many-body orders.
The balancing point corresponding to many-body-order-dependent basis sizes is marked with black circles. 
}
\label{fig:BalancingWithMixing}
\end{figure}

In order to formalize this more refined approach, we start by computing the basis-size uncertainties of the order-$p$ contribution in analogy to \cref{eq:MSError}.
They are denoted as $\AcademicMSErrorComp{p}(\emax)$ and displayed in \cref{fig:BalancingWithMixing} for the calculation of \elm{Ni}{78} discussed above.
The slower convergence of $\EnergyCorrection{2}$ compared to $\HFEnergy$ and $\EnergyCorrection{3}$ is reflected in the uncertainties, which are (over a large range in $\emax$) larger for the second-order contribution.
Of particular relevance is the finding that
\begin{equation}\label{eq:23unc}
    \AcademicMSErrorComp{2}(\emax) > \AcademicMSErrorComp{3}(\emax) \,,
\end{equation}
which is found to be robust for different nuclei and for both the \MagicInt and the \DoublyMagicInt~\cite{Arthuis2026DoublyMagic} Hamiltonian.
The different convergence rates can be used to further save computational resources in MBPT(3) calculations:
one simply must compute $\EnergyCorrection{3}$ at a smaller $\emax$ than $\EnergyCorrection{2}$. More specifically, this must be done by computing the two corrections in two separate calculations:
First, a HF calculation and a calculation of the second-order correction on top of it. 
Second, a HF calculation using a smaller basis and a calculation of the more costly third-order correction on top of that.
Such a calculation is said to be performed based on (many-body-)\emph{order-dependent basis sizes}. 
This strategy contrasts with the traditional approach where all orders are computed using the same one-body basis size and hence require only a single HF calculation.
Calculating the costliest many-body contribution in a smaller basis leads to a potentially significant computational advantage.

A calculation of this type combines values obtained from different basis sizes.
We label the $\emax$ associated with the $p$th-order contribution as $\emaxComp{p}$, with $p=1$ assigned to the reference-state energy,
\ie, $\emaxComp{1} = \emax^\text{HF}$.
To keep the notation compact, the vector $\emaxvec \equiv (\emaxComp{1}, \dots, \emaxComp{P})$ is employed. The ground-state energy is then computed according to
\begin{equation}\label{eq:energy_emax_p}
    \MBPTEnergy{P}(\emaxvec) = \HFEnergy(\emax^\text{HF}) + \sum_{p=2}^P \EnergyCorrection{p}(\emaxComp{p}) \,,
\end{equation}
where $\emaxComp{p}$ refers both to the basis size employed to compute the $p$th contribution and to the basis size used to build the reference state for that particular calculation. 
Finally, the total basis-size uncertainty of $\MBPTEnergy{P}(\emaxvec)$ is computed in analogy to Eq.~\eqref{eq:MSError}, \ie,
\begin{equation}
    \AcademicMSError{P}(\emaxvec) 
     = \Biggl|\sum_{p=1}^P \Delta \EnergyCorrection{p}(\emaxComp{p}) \Biggr| \,.
    \label{eq:TotalModelSpaceError}
\end{equation}
Having defined how to combine calculations with different one-body bases into a single value for the ground-state energy and its uncertainty, it remains to specify how to actually determine the one-body bases to be combined.

\subsection{Finding optimal basis sizes}\label{sec:mix_algo}

An iterative algorithm to find the optimal order-dependent basis sizes is now detailed. 
The procedure is initialized from a small one-body basis size and iteratively increases $\emax$. 
As an increase of $\emax$ by two units strongly increases the computational cost, the total resource consumption is dominated by the final computations.
The main task consists of choosing, at every step, the contribution for which the basis size
needs to be enlarged. Here, a ``greedy" update is devised such that $\emaxComp{p}$ for the order $p$ associated with the largest basis-size uncertainty is increased. The procedure is formalized as follows.
\begin{enumerate}
    \item \textbf{Initialization:} Compute the observable of interest at a small $\emax$ and evaluate its many-body truncation and basis-size uncertainties for all contributions up to order $P$.
    
    \item \textbf{Reduce individually large errors:} For each $p$, check whether $\AcademicMSErrorComp{p}(\emaxComp{p}) \geqslant \AcademicMBError{P}(\emaxvec)$.
    If yes, increase $\emaxComp{p}$ by two units and recompute the uncertainties. 
    Always set $\emaxComp{1} = \mathrm{max}(\emaxvec)$.%
    \footnote{
        As every energy correction is computed on top of a reference state constructed using the same one-body basis, $\HFEnergy$ is necessarily available for $\emaxComp{1} = \mathrm{max}(\emaxvec)$.
        For other correlation-expansion methods than MBPT, a similar reasoning applies to any $p < P$.  
    }
    This step is repeated iteratively until $\AcademicMSErrorComp{p}(\emaxComp{p}) < \AcademicMBError{P}(\emaxvec)$ for all $p$.%
    \footnote{
        For $\AcademicMSErrorComp{p}(\emaxComp{p}) \geqslant \AcademicMBError{P}(\emaxvec)$ the condition $\AcademicMSError{P}(\emaxvec) < \AcademicMBError{P}(\emaxvec)$ could only be met because of cancellations.
        Given that uncertainty models are not exact, the true total uncertainty could be strongly underestimated in that case.
        Therefore, one wants to avoid such cancellations and increase all $\emaxComp{p}$ until the individual uncertainties are small enough.
    }

    \item \textbf{Suppress leading error:} If $\AcademicMSError{P}(\emaxvec) \geqslant \AcademicMBError{P}(\emaxvec)$, determine the order $p$ with the largest $\AcademicMSErrorComp{p}(\emaxComp{p})$.
    Increase $\emaxComp{p}$ corresponding to that $p$ by 2 units.
    Set $\emaxComp{1} = \mathrm{max}(\emaxvec)$.
    Recompute the uncertainties.
    This step is repeated iteratively.
    
    \item \textbf{Stopping criterion:} If $\AcademicMSError{P}(\emaxvec) < \AcademicMBError{P}(\emaxvec)$ after step 2 or after an iteration step of 3, the obtained values of $\emax$ are sufficient: $\emaxBalancingvec = \emaxvec$.
    The final result is computed from Eq.~\eqref{eq:energy_emax_p}.

    If $\AcademicMSError{P}(\emaxvec) \geqslant \AcademicMBError{P}(\emaxvec)$ even with the maximal computational resources available, the balancing criterion cannot be fulfilled.
\end{enumerate}
A detailed example of the algorithm using actual numerical data is provided in \cref{sec:algo_example}.

\section{Balancing across the nuclear chart}\label{sec:chart}

\subsection{Uniform basis size}\label{sec:traditional}

In this section, the prescription to balance basis-size and many-body truncation uncertainties is validated numerically for a selected set of closed-shell nuclei ranging from \elm{O}{16} to \elm{Pb}{208}.
Two different sets of chiral two- and three-nucleon interactions are employed: the \MagicInt{} Hamiltonian from Ref.~\cite{Hebeler2011Magic}, for which $\OrderRatio{3}=0.44$  is used as explained in \cref{sec:mb}, and the \DoublyMagicInt Hamiltonian from Ref.~\cite{Arthuis2026DoublyMagic}, for which $\OrderRatio{3}=0.14$ is used based on a MBPT(4) calculation of \elm{Ca}{48}~\cite{Li2026HighOrderMBPT,LiPrivComm}.
The Hamiltonian matrix elements are generated with the \texttt{NuHamil} code~\cite{Miyagi2023NuHamil}. 
As before, $\emaxmax=16$ is employed, except for \elm{Pb}{208} where $\emaxmax=18$ is used for contributions up to second order.
Spherical MBPT calculations are performed with the \texttt{IMSRG++} code~\cite{StrobergIMSRG++}.
Three-nucleon interactions are included using the normal-ordered two-body approximation with an additional cut $e_1 + e_2+e_3 \leqslant \eThreeMax=24$ to reduce the number of three-body matrix elements~\cite{Miyagi2022HeavyAbInitio}. 
The sHO frequencies are chosen such that optimal convergence of the ground-state energy is obtained; see \cref{tab:freq} for their respective values.

\begin{table}[tbp]    
    \caption{sHO frequencies used in this work (in MeV).}
    \label{tab:freq}  
    \begin{ruledtabular}
    \begin{tabular}{lcccccc}
      & \elm{O}{16} & \elm{Ca}{40} & \elm{Ni}{78} & \elm{Sn}{100} & \elm{Sn}{132} & \elm{Pb}{208} \\
    \midrule
    \MagicInt & 20 & 18 & 14 & 16 & 14 & 12 \\
    \DoublyMagicInt & 18 & 16 & 14 & 16 & 14 & 12 \\
    \end{tabular}
    \end{ruledtabular}
\end{table}

\begin{figure}[bp]
\includegraphics[width=.9\linewidth]{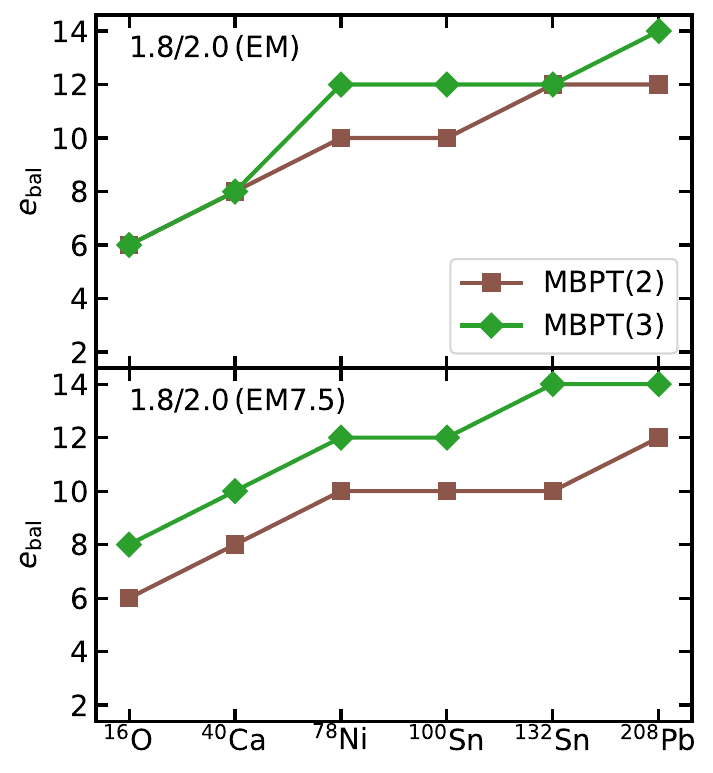}
\caption{\label{tab:balancing}Balancing points of ground-state energies of doubly closed-shell nuclei computed from the \MagicInt (top panel) and \DoublyMagicInt (bottom panel) Hamiltonians at the MBPT(2) and MBPT(3) levels.}    
\end{figure}

An overview of the results obtained using a uniform basis size is presented in \cref{tab:balancing}.
The balancing points increase smoothly with mass for both Hamiltonians, from $\emax=6/8$ in \elm{O}{16} to $\emax=12/14$ in \elm{Pb}{208}. This is consistent with the empirical fact that calculations of heavier nuclei require larger one-body basis sizes. In agreement with the balancing principle, the more accurate MBPT(3) truncation typically demands larger bases than MBPT(2). 
However, for the \MagicInt Hamiltonian, this does not always hold.
This is a consequence of the relatively large value of $\OrderRatio{3}$ employed for this Hamiltonian, which makes the many-body truncation uncertainty similar in magnitude at second and third order.

\begin{figure}[tbp]
\includegraphics[width=.9\linewidth]{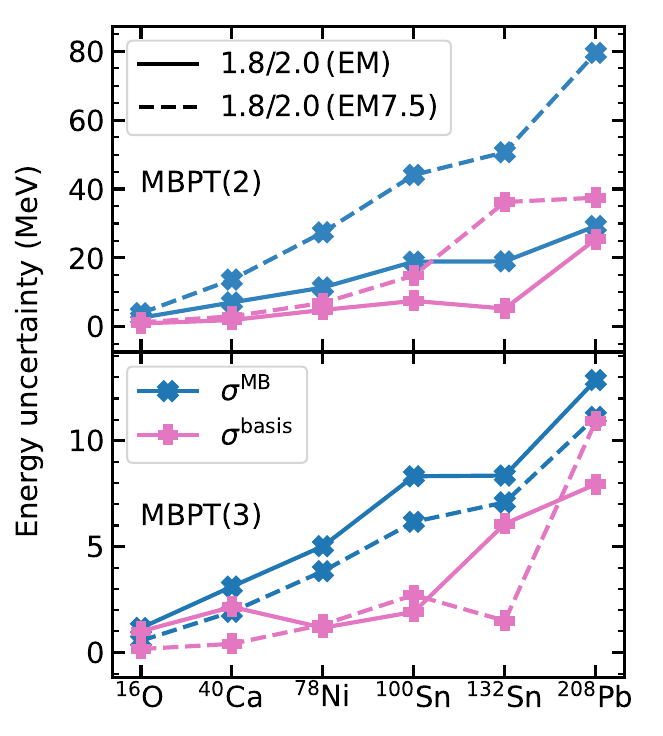}
\caption{\label{fig:uncertainties}Many-body truncation and basis-size uncertainties of ground-state energies of several doubly closed-shell nuclei at the balancing points from \cref{tab:balancing}.} 
\end{figure}

Let us now compare the balancing points obtained with the two Hamiltonians.
While they are very similar at second order, the balancing points are smaller for \MagicInt than for \DoublyMagicInt in half of the cases at third order. 
Let us unpack this with the help of \cref{fig:uncertainties}, where the uncertainties at the balancing points are displayed for the set of doubly-magic nuclei.
The top panel demonstrates that MBPT(2) calculations have a much smaller many-body truncation uncertainty for the \MagicInt Hamiltonian than for \DoublyMagicInt one.
This is in agreement with the fact that \MagicInt is known to be softer than \DoublyMagicInt.
The smaller many-body truncation uncertainty sets a more demanding threshold in the balancing procedure, but this is counteracted by the basis-size convergence, which is slower for \DoublyMagicInt.
The nontrivial interplay between these two factors turns out to lead to essentially identical balancing points for both Hamiltonians at second order.
This changes at the MBPT(3) level because the employed $\OrderRatio{3}$ is much smaller for \DoublyMagicInt than for \MagicInt, making the MBPT(3) many-body truncation uncertainties similar for both Hamiltonians (bottom panel).
Thus, the slower basis-size convergence for \DoublyMagicInt requires larger one-body bases to make the basis-size uncertainty match the MBPT(3) many-body truncation uncertainty.

The average relative many-body truncation uncertainty of the ground-state energies of the nuclei studied is equal to 1.9\% at second order and to 0.8\% at third order for the \MagicInt Hamiltonian. 
For the \DoublyMagicInt Hamiltonian, the averages read 4.3\% and 0.6\%, respectively.
These ratios mark the intrinsic accuracy of MBPT(2/3) calculations and are relatively stable across various nuclei as expected from size extensivity. 
\Cref{fig:uncertainties} also shows the basis-size uncertainties.
They are always smaller than the many-body truncation uncertainties by construction and their ratio to the energy is found to be less stable.

\subsection{Order-dependent basis sizes}\label{sec:mix_chart}

\begin{figure}[tbp]
\includegraphics[width=.9\linewidth]{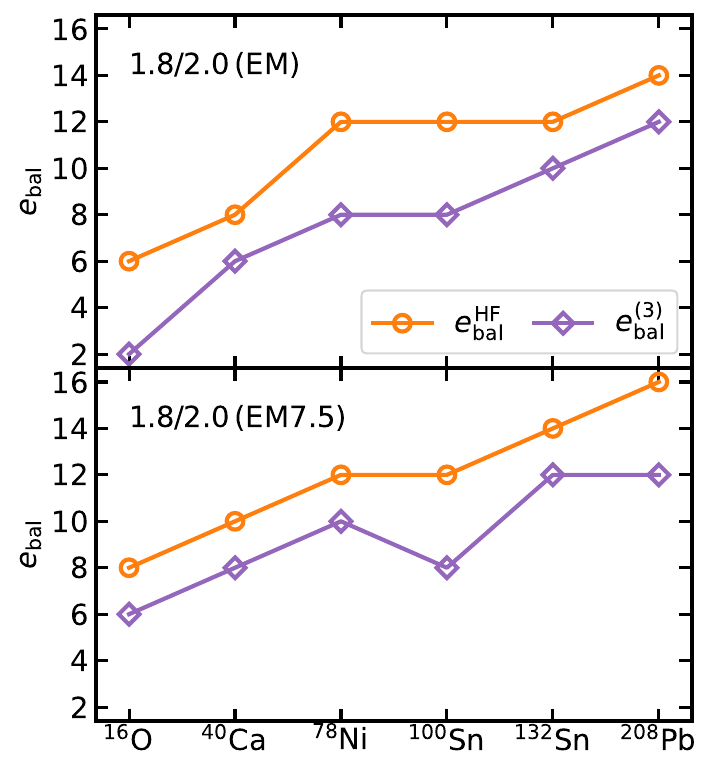}
\caption{
\label{tab:simple_mixing} Same as \cref{tab:balancing} for balancing points obtained at the MBPT(3) level with order-dependent basis sizes.
In all cases $\emaxBalancingHF = \emaxBalancingComp{2}$, except for \elm{Sn}{132} with \DoublyMagicInt, where $\emaxBalancingComp{2}=12$.
}
\end{figure}

Let us now investigate order-dependent basis sizes in MBPT(3) calculations.
To this end, the balancing algorithm described in \cref{sec:mix_algo} is applied to the same six nuclei as before.
The resulting balancing points are depicted in \cref{tab:simple_mixing}, where for both Hamiltonians it is observed that, in almost all nuclei,
\begin{equation}
    \emaxBalancingComp{3} < \emaxBalancingComp{2} = \emaxBalancingHF \,.
\end{equation}
The only exception is \elm{Sn}{132} for the \DoublyMagicInt Hamiltonian, where $\emaxBalancingComp{3} = \emaxBalancingComp{2} < \emaxBalancingHF$.
This general behavior is a consequence of the faster convergence with $\emax$ of the third-order energy correction compared to the lower-order contributions. This constitutes a key empirical observation given that the third-order contribution dominates the cost of MBPT(3) calculations.

It is now of interest to compare the MBPT(3) balancing points obtained for uniform basis sizes (see \cref{tab:balancing}) with those obtained here using order-dependent basis sizes. In almost all cases, the order-dependent balancing points for $\HFEnergy$ and $\EnergyCorrection{2}$ are identical to those obtained for uniform basis sizes.
Only once, the (inexpensive) HF contribution needs to be computed with a larger one-body basis than in the uniform case.
In addition, $\emaxBalancingComp{3}$ is smaller than the uniform balancing point throughout.
These results show that a significant cost reduction can be achieved consistently without compromising the accuracy of calculations by computing different contributions in separately adapted one-body bases.

\subsection{Application to deformed systems}
\label{sec:deformed}

Still relying on MBPT calculations, the order-dependent balancing scheme is now tested for doubly open-shell nuclei.
We calculate the ground-state energy of \elm{Ti}{46} and \elm{Cr}{58} up to third order based on the \MagicInt Hamiltonian.
The three-body part of the interaction is approximated by a nucleus-dependent two-body interaction via a rank-reduction technique~\cite{Frosini2021InMedium3NReduction}.
The deformed HF solution that corresponds to the absolute minimum of the total-energy curve as a function of the axial quadrupole deformation is taken as the reference state. 
In both nuclei, this absolute minimum is prolate and carries an axial deformation parameter of $\beta_2 \approx 0.2$.

\begin{figure}[tbp]
\includegraphics[width=.9\linewidth]{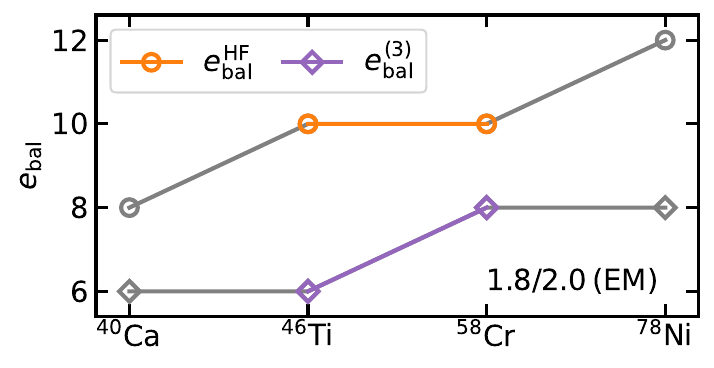}
\caption{
\label{fig:deformed}Same as \cref{tab:simple_mixing} with the balancing points  shown for two open-shell nuclei compared to those obtained for two closed-shell nuclei of similar mass (in gray).
In all cases $\emaxBalancingHF = \emaxBalancingComp{2} = \emaxBalancing$.}
\end{figure}

The extracted balancing points align well with the values extracted in doubly closed-shell isotopes in the same mass regime; see \cref{fig:deformed}. 
This indicates that the balancing points themselves are predominantly driven by size extensivity independently of the detailed characteristics of the reference state, \eg{}, of the presence of strong quadrupole collectivity.
In agreement with the previous findings in spherical closed-shell nuclei, third-order energy corrections can be evaluated using significantly smaller one-body bases, while the balancing points for the lower-order contributions are the same as obtained for a uniform basis size.
Given that deformed many-body computations are substantially more expensive than spherical ones, the reduction of the one-body basis size is even more advantageous here than for closed-shell systems.

\section{Coupled-cluster results}\label{sec:CC}

\subsection{Wave-function ansatz and uncertainties}\label{sec:CC_explanation}

Given that the idea of balancing theory uncertainties is generically applicable, it is presently extended to non-perturbative coupled-cluster (CC) calculations.
As the coupled-cluster expansion resums certain sets of perturbative corrections to infinite order, it often leads to more accurate results than MBPT calculations.
However, this comes at the price of significantly increased computing time and memory cost due to the need to solve a large system of coupled algebraic equations iteratively.
For reviews of CC theory, the reader is referred to Refs.~\cite{Hagen2014CCReview,ShavittBartlett2009Book}.

As for any  correlation-expansion method, the exact many-body ground state is accessed  by systematically adding dynamical corrections on top of an appropriate reference state $\ket{\Phi}$, here a spherical Hartree-Fock state.
In CC theory, the expansion relies on an exponential ansatz of the wave operator connecting the reference state to the exact ground state according to
\begin{align}
    |\Psi \rangle  \equiv e^T | \Phi \rangle \, ,
\end{align}
where the connected cluster operator $T\equiv T_1 + T_2+ ... + T_A$ involves a sum of $k$-particle-$k$-hole excitation operators $T_k$.
The CC hierarchy is defined by including higher-body operators in the definition of $T$, leading to the canonical CCSD ($T_{\text{CCSD}}\equiv T_1 + T_2$) and  CCSDT ($T_{\text{CCSDT}}\equiv T_1 + T_2 + T_3$) many-body truncations.
In practice, the full CCSDT solution is out of scope for realistic systems such that controlled approximations of it are employed. In this work, the CCSD[T] approximation, which captures the bulk of triples correlations at tractable computational cost, is used~\cite{Urban1985CCSDBracketsT}.

In analogy to \cref{eq:TruncatedMBPT}, the total CCSD([T]) energy can be written as
\begin{equation}
    E_\text{CCSD([T])} = \HFEnergy + E^\text{SD} (+ E^\text{[T]}) \,,
\end{equation}
where the three terms on the right-hand side denote the \emph{contributions} from HF, single\,+\,double excitations, and approximate triples, respectively.
The many-body truncation and basis-size uncertainties are estimated using the same models as for MBPT, but using $\emaxmax=14$ for the expensive triples calculations.
In quantum chemistry, 
\begin{equation}
    \CCOrderRatio{[T]} \approx \left| \frac{\ExactEnergy - E_\text{CCSD[T]}}{E^\text{[T]}} \right|
\end{equation}
is typically obtained to be around 0.2 for CCSD[T]~\cite{Noga1987CCSDT,Watts1990CCofBe,Watts1992CCofC2,Piecuch1996ApproximateT4}, suggesting that CCSD[T] captures around 98\% of the correlation energy.
This should not be confused with the often employed estimate $\CCOrderRatio{T} \approx 0.1$, which is valid for the much more expensive full CCSDT truncation~\cite{Bartlett2007CCChemistry}.
Because $\CCOrderRatio{[T]}$ depends on the Hamiltonian, it is important to consider the convergence of the CC expansion in nuclear systems.
A value of $\CCOrderRatio{T} \approx 0.1$ was obtained in \elm{He}{4}~\cite{Hagen2010CC} using the $\Lambda$-CCSD(T) variant that is more accurate than its simpler CCSD[T] counterpart. 
However, this number was obtained for a chiral Hamiltonian without 3N interactions.
Furthermore, \elm{He}{4} being a very light nucleus,  finite-size effects might be too large to use this estimate for heavier nuclei.
Estimating $\CCOrderRatio{T}$ in heavier systems is difficult because errors arising from truncating the CC expansion to finite order cannot easily be disentangled from other errors~\cite{Hergert2013IMSRG3N,Hergert2013Oxygen}.
Nevertheless one can note that $\CCOrderRatio{T} \approx 0.1$ is in agreement with what is obtained in electronic systems.
The same applies to $\CCOrderRatio{SD} \approx 0.1$ that is typically found in nuclei~\cite{Hagen2014CCReview, Sun2022RenormalizeCC, Vernik2026}.
Although better estimations are required in the future, we conclude that the quantum-chemistry value $\CCOrderRatio{[T]} = 0.2$ is likely approximately valid in nuclear systems and is thus employed here.

\begin{figure}[tbp]
\includegraphics[width=.9\linewidth]{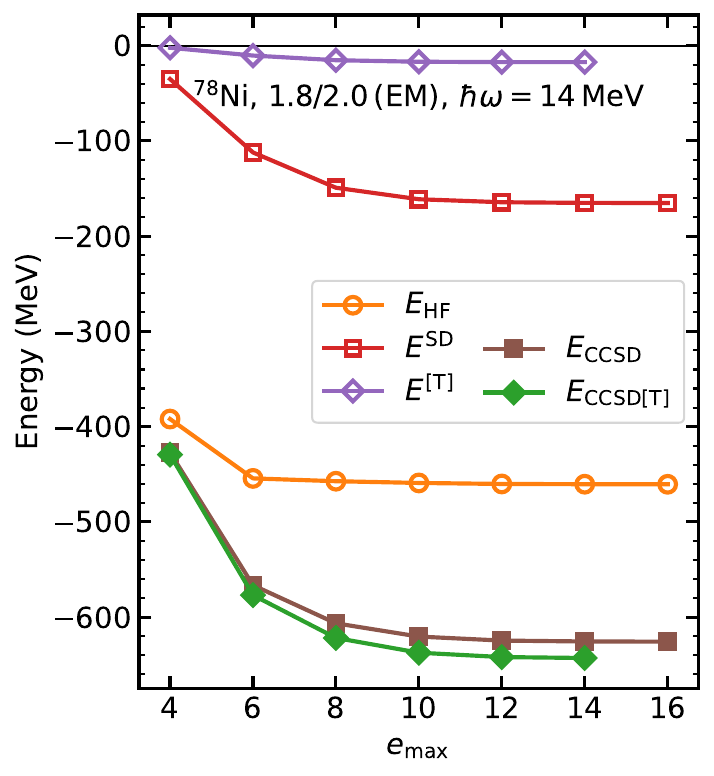}
\caption{Same as \cref{fig:convergence} for CC.}
\label{fig:CC_energies}
\end{figure}

\subsection{Balancing the coupled-cluster method}
\label{sec:CC_balancing}

The representative nucleus \elm{Ni}{78} is now used to demonstrate the assessment of theory uncertainties.
All calculations use the \MagicInt{} Hamiltonian and a sHO frequency $\freq= 14\MeV$. 
For previous CC calculations of \elm{Ni}{78} see Refs.~\cite{Hagen2016CC78Ni,Hu2024CC78NiTo70Ca}.
In \cref{fig:CC_energies}, the total ground-state energy and its individual contributions are displayed as a function of $\emax$.
The overall picture qualitatively agrees with the MBPT results from \cref{fig:convergence}.
In particular, triples corrections converge at a faster rate than the CCSD energy, thus motivating the use of many-body-order-dependent basis sizes for CC as well. Similar observations have been reported in quantum chemistry; see Refs.~\cite{Helgaker1997BasisSetConvergence,Martin2022BasisSetConvergence}.

\begin{figure}[tbp]
\includegraphics[width=0.9\linewidth]{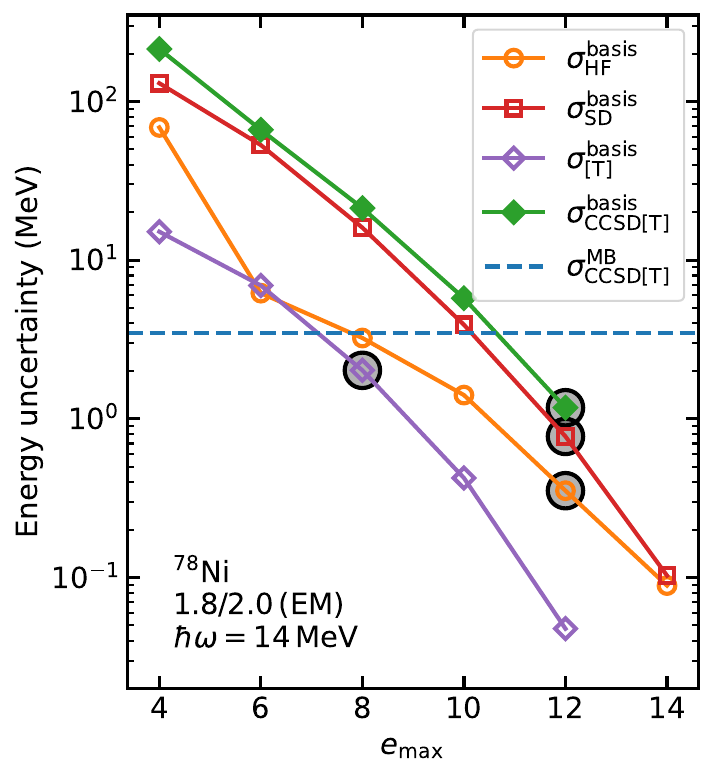}
\caption{Same as \cref{fig:BalancingWithMixing} for CCSD[T].
The basis-size uncertainty and the balancing point are also shown for the case where a uniform basis size is used.
}
\label{fig:CC_unc}
\end{figure}

Figure~\ref{fig:CC_unc} shows the uncertainties and the resulting balancing points at the CCSD[T] level. 
Using a uniform basis size, the balancing point is found to be $\emaxBalancing = 12$.
Using order-dependent basis sizes, the balancing point is rather $\emaxBalancingvec = (12, 12, 8)$, allowing triples to be computed with a significantly smaller basis than CCSD.
For the employed (soft) Hamiltonian, these balancing points agree with the ones obtained for MBPT(3), and the overall finding for CC aligns with the one based on MBPT calculations.
Clearly, the balancing principle is valid for CC (given an appropriate estimate of the many-body truncation uncertainty) and the algorithm for finding the balancing points can also be applied here. Using different $\emax$ for different contributions is particularly beneficial here due to the tremendous cost of computing triples corrections in CC for large one-body bases.

\section{A practical recipe}
\label{sec:rec}

The principle of balanced uncertainties allows one to identify optimal one-body basis dimensions for a given approximate many-body method.
However, so far theoretical uncertainties have been estimated using calculations based on large $\emaxmax$ values.
Clearly, this contradicts the idea of balancing uncertainties to avoid using too large one-body bases in the first place. 
This calls for finding pragmatic heuristics instead.

Rather than assessing the many-body truncation uncertainty by evaluating \cref{eq:MBError} at $\emaxmax$, it can be estimated for the $\emax$ of interest through
\begin{equation}\label{eq:MBErrorRunning}
    \PracticalMBError{P}(\emax) \equiv \OrderRatio{P} |\EnergyCorrection{P}(\emax)| \,.
\end{equation}
Estimating the basis-size uncertainty through \cref{eq:MSError} also requires calculations at $\emaxmax$.
We make the reasonable assumption that, at a given $\emax$, the finite basis-size error is dominated by the very next basis states left out when ordering the basis as a function of $e$. 
Together with the observation that the basis-size uncertainty falls off roughly exponentially with $\emax$, the following approximate uncertainty estimate is obtained:
\begin{align}\label{eq:MSErrorExtrapol}
    &\PracticalMSError{P}(\emax) \notag  \\
    & \equiv \frac{|\MBPTEnergy{P}(\emax - 2) - \MBPTEnergy{P}(\emax)|^2}{|\MBPTEnergy{P}(\emax - 4) - \MBPTEnergy{P}(\emax - 2)|} \,. 
\end{align}

\begin{figure}[tbp]
\centering
\includegraphics[width=.9\linewidth]{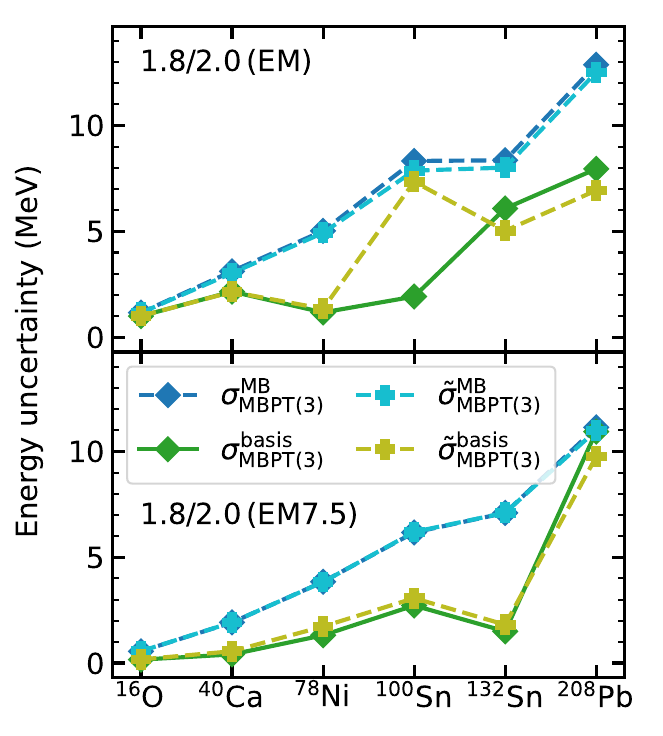}
\caption{Many-body truncation and basis-size uncertainties of ground-state energies at the balancing points, each estimated in two different ways: with \cref{eq:MBError,eq:MBErrorRunning} and with \cref{eq:MSError,eq:MSErrorExtrapol}, respectively.}
\label{fig:UncComp}
\end{figure}

Estimating the uncertainties with \cref{eq:MBErrorRunning,eq:MSErrorExtrapol} for a uniform basis-size scheme, the same balancing points as in \cref{sec:traditional} are found in almost all cases.
In addition, the uncertainties at the balancing points are very close to our earlier estimates; see \cref{fig:UncComp}.
The only exception is \elm{Sn}{100} for the \MagicInt Hamiltonian, where the presently obtained balancing point is $\emaxBalancing=10$ instead of 12, thus leading to significantly different basis-size uncertainties. 
Overall, the estimates \cref{eq:MBErrorRunning,eq:MSErrorExtrapol} are of sufficient quality to efficiently identify the balancing points.
In \cref{sec:practical_details}, different features of these uncertainty models are discussed before studying them in the context of order-dependent basis sizes, where their performance is slightly worse, yet sufficiently reasonable in practice.

Eventually, computing the uncertainty estimates at a given $\emax$ requires calculations with basis sizes no larger than the one associated with that $\emax$. This makes the balancing approach a tool of practical interest.

Based on our extensive benchmark using the \MagicInt{} interaction in \cref{sec:chart}, the following empirical recommendations for the evaluation of nuclear ground-state energies via MBPT and CC calculations can be made.
First, computational resources can be saved by computing higher-order corrections in smaller bases than lower orders.
This requires combining results from separate, but cheaper calculations as described in \cref{sec:mix_idea}.
\emph{In particular, the $\emax$ value needed to compute third-order corrections is 2 to 4 units smaller than the one required to compute the leading terms, which leads to significant computational savings.}
Second, since the balancing points seem to vary smoothly with mass number, the results obtained in \cref{sec:chart} can be used as guidance to perform MBPT(3) calculations of other nuclei.
At the optimal HO frequency, the following basis sizes are appropriate to compute the HF energy and the second-order correction:
\begin{itemize}
    \item \textbf{Light nuclei} ($A \lesssim 40$): For (medium-)light systems up to the $sd$-shell, particle-hole correlations converge quickly as a function of $\emax$ due to the location of the Fermi surface at low $e$ values.
    Hence, calculations for soft chiral interactions can be performed in $\emax \approx 6-8$ without compromising the accuracy of the many-body method. 
    \item \textbf{Medium mass} ($ 40 \lesssim A \lesssim 100$): In medium-mass systems up to the $pfg$-shell including neutron-deficient tin isotopes calculations can be performed with $\emax \approx 10-12$. 
    \item \textbf{Heavy nuclei} ($ 100 \lesssim A$): In heavy systems, the Fermi surface shifts towards higher shells thereby increasing the minimum basis size  that accommodates all nucleons. 
    Thus, calculations must be performed using one-body bases characterized by $\emax \approx 12-14$.
    For very heavy nuclei beyond \elm{Pb}{208}, even larger bases are expected to be required.
\end{itemize}
It must be emphasized that the above suggestions reflect empirical findings based on the \MagicInt{} Hamiltonian and must be revisited once other chiral Hamiltonians are employed. 
The same holds when using other many-body methods than MBPT or when going to higher orders, as this sets a more stringent accuracy goal.
For the soft \MagicInt interaction however, it seems that the recommendations given here also apply to CCSD[T] calculations.
Note also that nuclei close to the driplines require special care regarding the basis choice as continuum effects become important.

\section{Summary and outlook}\label{sec:summary}    

Nuclear \ai{} calculations are advancing towards heavy and exotic systems at ever-increasing accuracy.
This progress is accompanied by a steep increase in the demands for computational resources.
Therefore, it is mandatory to tame down the computational complexity of such calculations without compromising the quality of the results.
In this context, the present work emphasizes the importance of balancing many-body truncation and basis-size uncertainties based on the following simple guiding principle: 
\emph{Choose the smallest one-body basis that leads to a basis-size uncertainty smaller than the many-body truncation uncertainty.}  

In applying this rationale, third-order contributions at play in MBPT and CC high-accuracy calculations are shown to converge at a significantly faster rate with respect to the basis dimension than the lower orders. This finding is robust and has been validated in different systems and for different interactions. It can be exploited to combine low-order calculations in large enough basis sizes with higher-order calculations performed in smaller spaces.
Without deteriorating the predictive power of \ai{} calculations, this principle leads to a significant reduction of the computational burden.
Therefore, the flexibility to use many-body-order-dependent basis sizes defines a new path to accurate calculations of nuclear properties at reduced computational cost.
In addition to defining suitable models to estimate many-body truncation and basis-size uncertainties, we developed an algorithmic procedure to identify the balancing point in practical calculations. The procedure was validated based on explicit calculations of a set of representative nuclei.

The balancing procedure is generic and can be applied to any nucleus, correlation-expansion method, interaction, and observable, provided that robust uncertainty models are available.
For observables other than ground-state energies this is particularly challenging as size extensivity can often not be used to determine the associated many-body truncation uncertainty. 
Establishing appropriate uncertainty models is an important outstanding task for the nuclear theory community.

The treatment of basis-size uncertainties can be improved in complementary ways.
It is well known that convergence behavior can be drastically improved by employing optimized basis sets as demonstrated in quantum chemistry~\cite {Hehre1972PopleBasis,Dunning1989CorrelationConsistentBasis,Woon1995CorrelationConsistentBasisCore,Weigend2005AhlrichsBasis,Roos2005RelativisticANOBasisSets,Pritchard2019BasisSetExchange} and, within the context of natural orbitals, also in nuclear physics~\cite{Tichai2019natncsm,Novario2020a,Hoppe2020nat,Fasano2022NaturalOrbitals,Scalesi:2024pfm,Knoll2025HypernucleiNaturalOrbitals}.
In addition, basis-size extrapolation can be employed to extract an estimate of the infinite-basis limit from $\emax{}$ sequences.
In this way, one could reach balanced uncertainties with even smaller one-body basis dimensions.
Combining this with order-dependent basis sizes requires extrapolating separately contributions from different orders~\cite{Karton2020BasisSetExtrapolation,Martin2022BasisSetConvergence}.

The principle of balancing uncertainties does generally apply to all theory uncertainties and thus naturally extends to additional sources of errors present in (nuclear) many-body applications.
Important examples are the truncation of three-body interaction matrix elements~\cite{Miyagi2022HeavyAbInitio}, normal-ordering approximations~\cite{Roth2012NO2B,Gebrerufael2016mrno,Ripoche2020,Frosini2021InMedium3NReduction}, or the growing field of low-rank tensor decompositions~\cite{Tich21SVDNN,Zhu2021,Tichai2024rsvd,Frosini2024SVDMBPT2,Frosini2024SVDMBPT2Ge}.
Eventually, interaction uncertainties, emerging from the SRG used to accelerate many-body convergence~\cite{Roth2014SRGof3N, Miyagi2022HeavyAbInitio} and from truncations of the chiral EFT expansion~\cite{Furnstahl2015Uncertainties,Melendez2019Uncertainties,Hu2022AbInitio208Pb,Plies2025UncertaintiesSVD,Millican2026ChiralConvergence}, must be treated on an equal footing.

\section*{Acknowledgements}

The authors thank Matthias Heinz, Zhen Li, Achim Schwenk, and Isak Svensson for useful discussions and Stavros Bofos for help with the deformed calculations.
This work was supported in part by the European Research Council (ERC) under the European Union's Horizon Europe research and innovation programme (Grant Agreement No.~101162059), by the LOEWE Top Professorship LOEWE/4a/519/05.00.002(0014)98 by the State of Hesse, and by Fonds de la Recherche Scientiﬁque (F.R.S.-FNRS, Belgium) under the MIS Project nr. 40028446. The authors gratefully acknowledge the Gauss Centre for Supercomputing e.V. (www.gauss-centre.eu) for funding this project by providing computing time through the John von Neumann Institute for Computing (NIC) on the GCS Supercomputer JUWELS at J\"ulich Supercomputing Centre (JSC).

\appendix

\section{Example of balancing algorithm}
\label[appendix]{sec:algo_example}

To illustrate the implementation of the algorithm introduced in \cref{sec:mix_algo} ``in action'', it is applied to MBPT calculations of \elm{Ni}{78} based on the  \MagicInt Hamiltonian.
In the following, the functioning of the algorithm is described step by step using order-dependent basis sizes.
The reader is invited to consult \cref{fig:BalancingWithMixing} displaying the uncertainties of each order, in parallel to the following description.
A summary of the description is provided in \cref{tab:algo_example}.

Step 1 of the algorithm is initialized at $\emax=4$.
The basis-size uncertainty is $\AcademicMSError{3}(\emaxvec) \approx 211 \MeV$ which is much larger than $\AcademicMBError{3} \approx 5.0 \MeV$, clearly indicating a too small basis size. 
Indeed, at this point $\AcademicMSErrorComp{p}(\emax) > \AcademicMBError{3}$ for every $p\leqslant P=3$.

\begin{table}
    \caption{Summary of the balancing algorithm applied to MBPT(3) calculations of \elm{Ni}{78} based on the  \MagicInt Hamiltonian. The threshold is set by the many-body truncation uncertainty $\AcademicMBError{3} \approx 5.0 \MeV$.
    }
    \label{tab:algo_example}
    \begin{ruledtabular}
    \begin{tabular}{ccccc}
        Step & $\emaxComp{1}$ & $\emaxComp{2}$ & $\emaxComp{3}$ & $\AcademicMSError{3}$ (MeV) \\
        \midrule
        1 & 4 & 4 & 4 & 211 \\
        2 & 6 & 6 & 6 & 65 \\
        2 & 8 & 8 & 6 & 23 \\
        2 & 10 & 10 & 6 & 9.7 \\
        3 & 10 & 10 & 8 & 7.0 \\
        3 & 12 & 12 & 8 & 3.1 \\
    \end{tabular}
    \end{ruledtabular}
\end{table}

The algorithm continues with step 2 increasing all components of $\emaxvec$ to 6, at which point $\AcademicMSErrorComp{3}(\emaxComp{3}) < \AcademicMBError{3}$. 
Thus, it remains to increase $\emaxComp{1}$ and $\emaxComp{2}$ to 8.
Now, the criterion is fulfilled for $p=1$, such that only $\emaxComp{2}$ is further increased to 10.
At this point, $\AcademicMSErrorComp{p}(\emax) > \AcademicMBError{3}$ for every $p$ and step 2 is finished by setting $\emaxComp{1} = \mathrm{max}(\emaxvec) = 10$. 

The current configuration reads $\emaxvec = (10, 10, 6)$ and the total basis-size uncertainty is now reduced to $\AcademicMSError{3}(\emaxvec) \approx 9.7 \MeV$. Since it is still greater than the many-body truncation uncertainty, step 3 is executed. 
At this point, the largest basis-size uncertainty is given by $\AcademicMSErrorComp{3}(\emaxComp{3})$.
Therefore, $\emaxComp{3}$ is increased to 8.
Now, $\AcademicMSError{3}(\emaxvec) \approx 7.0 \MeV$, which is still too large.
Thus, the iteration continues.
The new largest basis-size uncertainty is $\AcademicMSErrorComp{2}(\emaxComp{2})$, so $\emaxComp{2}$ is increased to 12.
This makes it necessary to also increase $\emaxComp{1}$ to the same value.

Finally, $\AcademicMSError{3}(\emaxvec) \approx 3.1 \MeV < 5.0 \MeV \approx \AcademicMBError{3}$.
Therefore, the balancing criterion is reached (step 4) with the final model-space configuration $\emaxBalancingvec = (12, 12, 8)$.
This balancing point is marked in \cref{fig:BalancingWithMixing}.
Note that no expensive calculations of third-order contributions at $\emaxComp{3}=12$ had to be performed, contrary to if all contributions had been computed using the same basis size, \cf\ \cref{fig:BasicBalancing}.

\section{Details on $\PracticalError$ uncertainty models}\label[appendix]{sec:practical_details}

Details are now provided regarding the uncertainty models introduced in \cref{sec:rec} that make it possible to actually obtain a computational advantage in practical calculations. 
This is again done using MBPT(3) calculations of \elm{Ni}{78} with the \MagicInt Hamiltonian.

\begin{figure}[tbp]
\centering
\includegraphics[width=.9\linewidth]{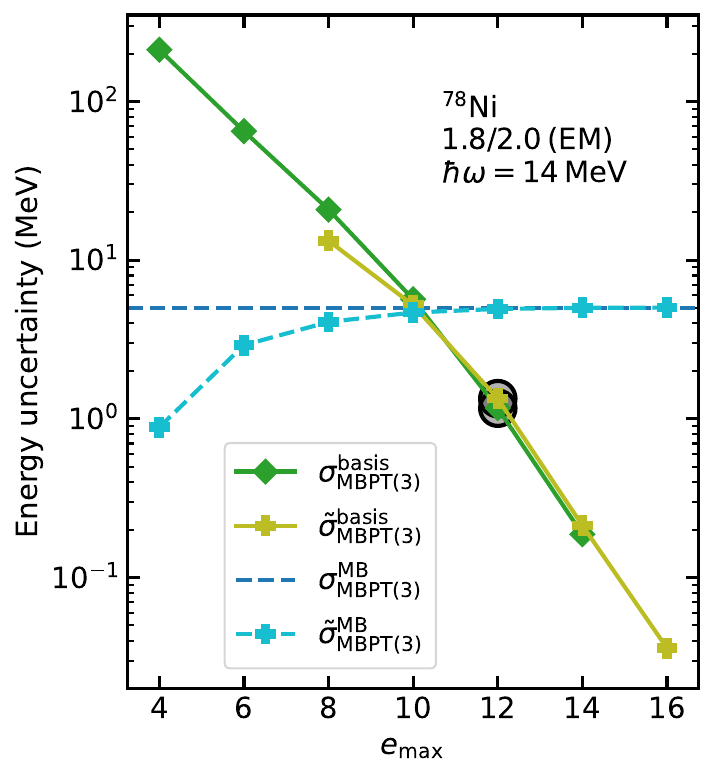}
\caption{Same as \cref{fig:BasicBalancing} using two different sets of many-body truncation and basis-size uncertainty models.}
\label{fig:PracticalBalancing}
\end{figure}

Let us start by comparing $\PraMBError$ [\cref{eq:MBErrorRunning}] with $\MBError$ [\cref{eq:MBError}] as a function of $\emax$.
The two many-body truncation uncertainty estimates are depicted in \cref{fig:PracticalBalancing} using dashed lines.
While the two estimates differ in small bases, they are very similar for large $\emax$ values, as expected.
Around the balancing point, the difference between both uncertainty estimates is very small such that $\PraMBError$ can be safely used in practice.

A similar conclusion applies to $\PraMSError$.
As can be seen from \cref{fig:PracticalBalancing}, it is close to $\MSError$ for all basis sizes where values are available.
Note that no values are shown for $\PraMSError$ at $\emax=4$ and~6.
This results from a slight drawback of this uncertainty model when applied at a given $\emax$ as it requires results from computations performed at three different one-body basis sizes, starting from $\emax\!-\!4$.
For small $\emax$, these values are simply not attainable because the model spaces are too small to even accommodate all nucleons. 
Because this problem occurs only at small $\emax$, finding the balancing point is not affected in any of the calculations performed with uniform basis sizes, \ie\ the resulting balancing points are robust.

\begin{figure}[tbp]
\includegraphics[width=.9\linewidth]{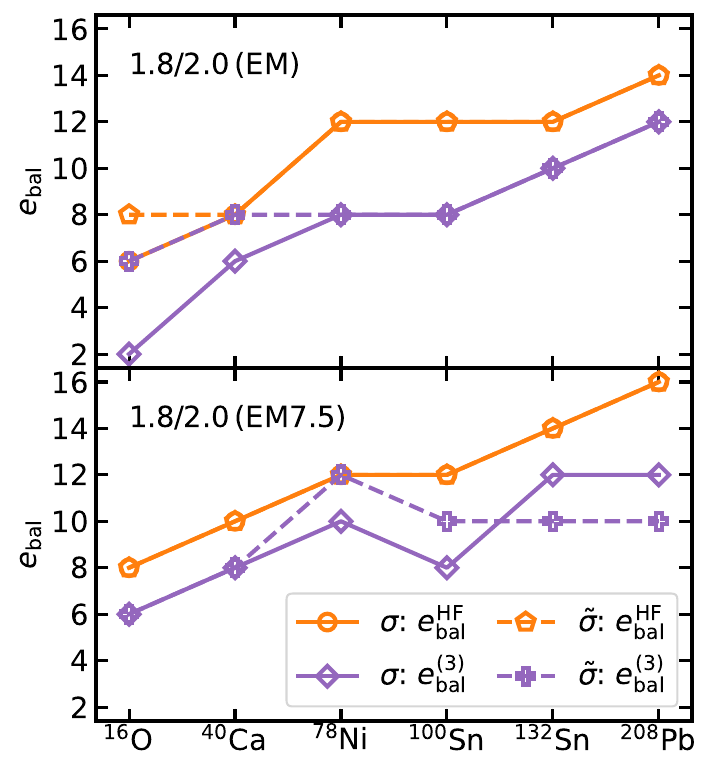}
\caption{Same as \cref{tab:simple_mixing} using two different sets of uncertainty models.
In all cases $\emaxBalancingHF = \emaxBalancingComp{2}$, except for \elm{Sn}{132} with \DoublyMagicInt when using $\sigma$, where $\emaxBalancingComp{2}=12$.
}
\label{fig:MixPractical}
\end{figure}

Let us now turn to finding order-dependent balancing points when applying the algorithm of \cref{sec:mix_algo} based on the $\PracticalError$ uncertainty models.
In \cref{fig:MixPractical}, the balancing points are compared with the ones obtained in \cref{sec:mix_chart} using the $\Error$ uncertainty models.
In half of the cases, the balancing points agree, but in the other half, at least one component of $\emaxBalancingvec$ differs slightly.
In most of those cases, the determined balancing points are too conservative when using $\PracticalError$.
Only once [\elm{Pb}{208} with \DoublyMagicInt] a configuration is found where the balancing criterion is fulfilled prematurely. For \elm{Sn}{132} with the same Hamiltonian, a different but equally valid balancing configuration is obtained.

There are two main reasons for these findings.
The first is related to the exponential extrapolation that underlies $\PraMSError$.
It works well for total ground-state energies because their basis-size uncertainties typically decay exponentially.
This is often less true for individual contributions as their convergence curves are somewhat less smooth.
Therefore, $\PraMSError$ performs a bit worse in those cases. 
Second, a drawback of $\PraMSError$ already mentioned above has negative consequences here:
Calculations at $\emax\!-\!4$ are impossible when $\emax$ is small, as the basis sizes become too small to even contain all nucleons.
This means that in a few cases, $\PracticalMSErrorComp{3}$ cannot be computed at the true balancing point known from \cref{tab:simple_mixing}. 
In other words, the principle of balanced uncertainties suffers from its own success:
it would limit calculations to basis sizes that are too small to prove that they are large enough.
In addition, $\PraMSError$ is not additive. 
Instead, the sum $\sum_{p=1}^P \PracticalMSErrorComp{p}$ often overestimates $\PracticalMSError{P}$.
Overall, these issues suggest that finding better basis-size uncertainty models for the  contributions originating from each many-body expansion order would be of interest.
Nevertheless, the balancing points obtained with $\PracticalError$ are in general reasonable (although not perfect), even for order-dependent basis sizes.

\bibliography{Literature}

\end{document}